\documentclass{article}
\PassOptionsToPackage{numbers, compress}{natbib}
\usepackage[preprint]{neurips_2026}

\usepackage{amssymb}
\usepackage{algorithm}
\usepackage{algpseudocode}
\usepackage{booktabs}
\usepackage{xcolor}
\usepackage{natbib}
\usepackage[table]{xcolor}  
\usepackage[table]{xcolor}
\definecolor{rowshade}{gray}{0.93}

\usepackage[utf8]{inputenc} 
\usepackage[T1]{fontenc}    
\usepackage{hyperref}       
\usepackage{url}            
\usepackage{booktabs}       
\usepackage{amsfonts}       
\usepackage{nicefrac}       
\usepackage{microtype}      
\usepackage{xcolor}         
\usepackage{graphicx}       
\usepackage{subcaption}     
\usepackage{multirow}       
\usepackage{enumitem}       
\usepackage{wrapfig}        
\usepackage{amsmath}

\usepackage{fontawesome}
\usepackage{float}
\usepackage{placeins}
\usepackage{longtable}
\usepackage{xspace}
\usepackage[capitalize]{cleveref}
\usepackage{todonotes}
\crefname{figure}{Fig.}{Figs.}
\Crefname{figure}{Fig.}{Figs.}
\crefname{section}{Sec.}{Secs.}
\Crefname{section}{Sec.}{Secs.}

\title{Language Models Embody and Amplify Human Cognitive Distortions: What is to be Done?}

\author{%
  Arnau Marin-Llobet \\
  Harvard University \\
  Cambridge, MA 02138 \\
  \And
  Steven A. Lehr \\
  Cangrade, Inc. \\
  Watertown, MA 02472 \\
  \And
  Mahzarin R. Banaji\\
  Department of Psychology \\
  Harvard University \\
  Cambridge, MA 02138 \\
  \texttt{mahzarin\_banaji@harvard.edu} \\
}

\begin{document}

\maketitle

\begin{abstract}
Human judgment is fundamentally prone to error. A promise of AI is that it will rid decisions of bias and ensure a fairer and safer world for all. Yet research unequivocally demonstrates that LLMs exhibit consequential sociocognitive biases. We alert readers that bias in AI (a) is covert and ironically a feature of alignment goals, (b) is not merely a mirror, but an amplifier of human bias, (c) intensifies across model generations, and (d) even transmits bias to humans. Given the potentially seismic and ubiquitous influence of AI on decision making, we propose countermeasures that are diagnostic, regulatory and operational.
\end{abstract}

\section{The Problem}

Few technologies in history have generated such fervid optimism as the newest frontier of AI, large language models (LLMs). And yet, for all its promise of advancing human progress, this technology has also provoked widespread fear and anger, including from those of opposing ideologies. Among the causes of this backlash is that bias in AI far exceeds tolerable levels and that developers exude a reckless disregard of this fact.

In this commentary, we aim to achieve two goals. First, we summarize a growing body of research showing that LLMs exhibit humanlike sociocognitive biases that are broadly held, deeply embedded, and often covert. Second, we propose countermeasures to safeguard against model bias including diagnostic, regulatory and operational solutions.

\section{The Evidence}

Human judgment is systematically and fundamentally prone to error. This observation, a cornerstone of experimental psychology, has been the basis of at least three Nobel Prizes and hundreds of respected scientific careers. Thousands of studies have shown that humans hold conscious prejudices and less conscious implicit biases in judgments involving social categories of gender, race, sexuality, body weight, disability, and countless other dimensions \cite{banaji2016blindspot}. Years before LLMs were publicly deployed, research on word embeddings showed that massive text corpora -- those that would form the substrate for LLMs -- reproduced each and every one of these human implicit associations. Even then, experts warned that these biases would be transmitted to AI models trained on language \cite{caliskan2017semantics}. Inexplicably, developers failed to adequately heed these warnings. One possible explanation for this is overconfidence (another immutable law of human psychology). The creators of AI models have been enamored with the idea that safety issues could be solved through technical processes for aligning AI models with human preferences and values. But this view overlooks the serious problem that the humans with whom the models are being aligned themselves fall prey to implicit biases.

It has now become clear that LLMs have indeed acquired the many biases of the species that created them. These arise for every consequential social category where tests have been conducted: LLMs show bias around race, skin tone, gender, age, disability, dialect, ethnicity, religion and weight, to name just a few \cite{bai2025explicitly, zhao2025explicit, hofmann2024ai}. These biases emerge easily, not as edge cases coaxed out by unfair prompts. They surface on ordinary requests, and they survive the very fine-tuning meant to remove them \cite{bai2025explicitly, zhao2025explicit, hofmann2024ai, panickssery2024llm, lehr2026face}. Nor are biases limited to these frequently scrutinized social categories. Asked to evaluate written work, LLMs rate their own outputs more highly than those of other models, even when human annotators judge the texts to be of equal quality \cite{panickssery2024llm}, a clear self-serving bias that can quietly corrupt the work of business, law, and science. LLMs even make biased judgments from face morphology alone, readily making selection decisions based on false inferences about traits like trust and competence \cite{lehr2026face}. Worse still, evidence suggests that most of these biases have persisted -- and indeed often worsened -- in later models \cite{bai2025explicitly, zhao2025explicit, hofmann2024ai, lehr2026face}.

This pattern extends to reasoning itself. On classic tasks from psychology, LLMs reproduce the catalogue of human characteristics and cognitive biases, frequently in exaggerated form relative to human data \cite{binz2023using, lehr2025kernels, cheung2025large}. Furthermore, models can be swayed by the same persuasion and propaganda tactics that work on gullible humans \cite{meincke2026persuading}. These findings highlights yet another kind of bias: factors like apparent authority, scarcity and affiliation can be effortlessly used to change model behavior. Beyond questions of fairness, this is a safety problem: a system that can be talked into harmful compliance can also be more easily aimed toward nefarious goals.

Some of the most consequential biases are also the best hidden. The same systems that describe African Americans in positive terms when asked directly will make covertly racist judgments when the only cue is a person's dialect \cite{hofmann2024ai}. Given identical content expressed in African American English compared to Standard American English, several widely used models were more likely to assign speakers to lower-status jobs, to convict them, and to sentence them to death; the stereotypes driving these judgments were more negative than anything contemporary humans willingly say.

Perhaps more startling than the presence of these biases is the degree to which new generations of language models have mirrored the human implicit--explicit divide. Language models often pass standard explicit bias benchmarks with ease. Ask a model whether it prefers thin people over fat people, or white over Black people, and you will quickly be humbled into feeling ashamed for even asking. Yet when tested without explicit mention of social categories, models reveal pervasive biases spanning all the many domains in which we humans, ourselves, fail \cite{bai2025explicitly, zhao2025explicit, hofmann2024ai}. A painful irony is that one of the most foundational goals of LLM training, aligning models to human preferences, may itself be an instrument by which bias is propagated in models. Alignment training teaches a model to state the unbiased answer while leaving the biased association intact to drive its decisions, much as explicitly unbiased humans still display biased associations and behaviors. Indeed, while alignment training has been shown to reduce overt biases, it does not appear to impact covert biases, which thus continue to grow unencumbered as models scale \cite{zhao2025explicit, hofmann2024ai}. This fundamental oversight suggests that experts on social cognition, who might readily have predicted this problem, were not at the table to discuss the original approaches to model training.

A common reply to this litany of observed biases is that models merely mirror their training data, in which case the solution would simply be cleaner data. However, this argument is flawed for several reasons. First, these systems are not just mirrors but amplifiers of bias: studies typically find that effects in LLM outputs are exaggerated relative to the training data and the humans who produced it \cite{hofmann2024ai, lehr2026face, lehr2025kernels, cheung2025large, bianchi2023easily}. Second, AI models have been shown to introduce new biases that are not present in their training data at all. For example, LLMs show a distinct preference for LLM- over human-generated content, a bias unlikely to be traceable to AI training data \cite{laurito2025aiai}. Research also shows that in selection tasks, LLMs can develop emergent biases toward novel social groups, attributable to neither existing biases (since the groups are novel) nor ground truth (since the actual quality of the candidates was carefully controlled) \cite{wu2026large}. Finally, many LLM biases appear to be increasing with each new generation of models \cite{bai2025explicitly, zhao2025explicit, hofmann2024ai, lehr2026face, wu2026large}. In the face of such evidence, the goal of cleaner training data remains valuable, but it is unlikely to be sufficient to address what appears to be a more complex set of errors in model behavior.

To make matters worse, biases do not remain innocuously inside the model. Humans are deeply social creatures, and those working alongside AI tend to adopt model biases as their own \cite{wilson2025nothoughts}. In hiring experiments, participants paired with a moderately biased system mirrored its racial preferences almost exactly, choosing whichever group the LLM favored, and pushed back only when the bias was extreme and unmistakable. This closes a vicious loop: biased outputs shape human decisions, those decisions produce biased outcomes, and the outcomes flow back into the text and feedback data on which the next model is trained. Biases thus compound across generations, so that each new model embodies these distortions more deeply than the last. Ironically again, humans have become less biased over the centuries while these models, even in their first few iterations, are heading in the opposite direction.

\section{The Solution}

We write as scientists who believe in the promise of AI that is wisely executed. But we are alarmed by the scant attention paid to the rising evidence of model bias. The stakes are not abstract. These models are already being used to screen job applicants, provide medical advice, power lending decisions, and inform legal risk assessments, among other consequential domains. The sheer number of decisions these models will make -- with just a few LLMs already adopted by hundreds of millions of users -- means that their entrenched biases have the potential to corrupt human judgments on a previously unimaginable scale. The issue invites concern for public safety and the ideals of modern democracies, and as such, we must ask: what can be done? Specifically, we call for a transparent and evolving repository of LLM biases, with documentation of whether models are successively improving or degenerating. Drawing on regulatory frameworks from other industries, we then offer solutions to stimulate the development of less biased AI systems and a framework for cross-industry collaboration to address this growing public safety issue.

\paragraph{What human bias has taught us about debiasing.} As authors who have written about covert biases in humans, we lack the naivety to simply demand that models be debiased. While biases can be modulated in humans, to date, the goal of fully debiasing ourselves has proved intractable. Although this goal may be more feasible for machines that can be programmed relative to biological minds that evolved from natural selection, it is unrealistic to expect any simple method to entirely eradicate bias. However, humans and organizations have made progress around acknowledging and repairing biased decision processes. Success here has been tied to two key elements: transparency and proper alignment of incentives. To address biases, humans and organizations must first recognize and track them. And social, legal and economic mandates must nudge organizations toward accountability and action.

\paragraph{Public registration of cognitive and social biases in AI models.} Before any serious effort to debias models can be undertaken, detection and public reporting is imperative. This demand is analogous to safety requirements and testing routinely imposed on products, from car seats to cosmetics.

We recommend independent behavioral audits built on a shared, versioned battery of validated indirect bias measures, so that results are comparable across models and over time. Research shows that AI models, like humans, can express unequivocal disapproval of biases that nevertheless stealthily corrupt their decisions. For this reason, the proposed registry should focus on overt and covert bias, particularly in consequential domains. Fortunately, experimental psychology has spent a century building methods to do precisely this: detecting the gap between what an agent says and what that agent does.

This registry should be broad, avoiding the common pitfall of focusing only on biases like race and gender that are most prominent in public discourse. While these domains are unquestionably important, the ultimate goal should be to develop more broadly egalitarian AI systems, capable of avoiding bias even in instances where explicit guardrails and policies have not been engaged, an example being facial physiognomy bias. To enable developers to achieve this goal, the registry should cover a wide variety of sources of bias and in a variety of domains of application, from college admissions to medical diagnoses and legal decisions. Besides the public-serving function such a registry provides, its presence will benefit model developers, who will obtain concrete markers of model quality and avoid potential reputational damage and regular legal challenges.

\paragraph{Incentives and Public Mandates.} An AI bias registry without regulatory weight or consequences will be a mere academic exercise. To make tracking decision quality useful, incentives should be aligned to create an organizational mission of fairness. However, because these issues are complex and debiasing models is difficult, judiciousness will be required to achieve this without hampering progress with over-regulation. Here, lessons may be learned from other industries.

\textit{Incremental Standards (Home Appliances).} The Department of Energy does not expect appliance companies to achieve ambitious efficiency goals overnight. Instead, minimum standards for technologies like refrigerators and HVAC systems are routinely updated: manufacturers cannot release new models that consume more energy than these incrementally tightened baselines. Similarly, while AI developers cannot immediately eradicate bias, there is no reason for models to become more biased over successive generations.

\textit{Targeted Deployment Constraints (Pharmaceuticals).} Before a drug's release, companies must carefully test compounds for side effects. If they are found, the drug isn't inevitably prohibited outright, but its availability to vulnerable populations is limited. Similarly, when models are found to have a high level of bias, they could be barred from release in high-stakes domains.

\textit{Taxation Tied to Poor Performance (Automotives).} Automobiles with poor fuel efficiency are not banned outright but instead are assessed by the EPA. Manufacturers must pay a tax that scales with how poorly a car performs. Similarly, a bias tax could be levied upon model developers, with the amount depending on the degree of bias and the model's sales or usage.

\textit{Shared Liability (Construction).} If drywall in a new construction has toxic mold, the drywall manufacturer shares liability with the contractor who built the walls. Similarly, AI developers should share liability for bias with the models' users, and the developer's liability should be greater if it fails to disclose bias that makes model use risky or inappropriate.

\paragraph{Transparency, access, and collaboration.} While independent research scientists are hard at work testing models for bias, lack of transparency and full access to models render these efforts less efficient and effective than they should be. Nearly all the research outlined above was done by researchers outside the major AI companies, without access to model internals, training data, or deployment configurations. Other kinds of safety evaluators have begun to obtain deeper access; researchers studying bias have not, even though this domain is no less important. The same access should be granted to behavioral auditors: checkpoints from successive training stages, to separate the relative contributions of pretraining, fine-tuning and RLHF; retention (rather than deprecation) of older models, so behavior can be traced across generations; and disclosure of data sources, to help trace biases to their origins.

To address questions of the magnitude of this civilization transforming technology, more ideas will certainly be needed, and we call upon experts from economics, law, and public policy to contribute solutions that draw on the unique thinking of their fields. Moreover, no single sector can achieve any of this alone. Industry holds the models and data, academia specialized methods and expertise, and government the mandate and power for optimal regulation. Shared testbeds, pre-deployment review, and dedicated funding for replication would let each contribute what it does best.

In closing, we offer an inconvenient observation: despite the measured tone we have taken, these solutions may well prove laughably inadequate given the potential for bias in AI to radically undermine human fairness, dignity and wellbeing. We may collectively wish to signal the magnitude of the seismic shift we face by setting aside this focus on incremental rules and laws and frankly questioning the lack of decency -- decency, in the sense in which Joseph Welch challenged Senator McCarthy in 1954 -- with which this fast-emerging technology is being pursued.

Hundreds of millions of people now interact daily with systems that, on many measures, carry deeply embedded biases. The case for collective action is clear. It is time to work together to find solutions before irreparable harm is done.

\section*{Competing Interests}
A. Marin-Llobet and M. R. Banaji declare no competing interests. S. Lehr owns equity in Cangrade, a company involved in debiasing machine learning models. However, Cangrade does not build LLMs, did not fund this work, and is not expected to profit from its publication. A. Marin-Llobet is supported by Harvard Mind, Brain, Behavior Interfaculty Initiative, Coefficient Giving and the RCC-Harvard Fellowship. 

\FloatBarrier
{
    \small
    \bibliography{main.bib}
    \bibliographystyle{unsrt}
}
\FloatBarrier

\appendix
\counterwithin{figure}{section}
\counterwithin{table}{section}
\renewcommand{\thefigure}{A\arabic{figure}}
\renewcommand{\thetable}{A\arabic{table}}

\end{document}